# Drone-Based Multispectral Imaging and Deep Learning for Timely Detection of Branched Broomrape in Tomato Farms

Mohammadreza Narimani[a], Alireza Pourreza[*a], Ali Moghimi[a], Mohsen Mesgaran[b], Parastoo Farajpoor[a], Hamid Jafarbiglu[a]

[a] Department of Biological and Agricultural Engineering, University of California, Davis, CA, 95616, United States
[b] Department of Plant Sciences, University of California, Davis, CA, 95616, United States

## ABSTRACT

This study addresses the escalating threat of branched broomrape (Phelipanche ramosa) to California's tomato industry, which provides over 90% of the United States processing tomatoes. The parasite's life cycle, largely underground and therefore invisible until advanced infestation, presents a significant challenge to both detection and management. Conventional chemical control strategies, while widespread, are costly, environmentally detrimental, and often ineffective due to the parasite's subterranean nature and the indiscriminate nature of the treatments. Innovative strategies employing advanced remote sensing technologies were explored, integrating drone-based multispectral imagery with cutting-edge Long Short-Term Memory (LSTM) deep learning networks and utilizing Synthetic Minority Over-sampling Technique (SMOTE) to address the imbalance between healthy and diseased plant samples in the data. The research was conducted on a known broomrape-infested tomato farm in Woodland, Yolo County, California. Data were meticulously gathered across five key growth stages determined by growing degree days (GDD), with multispectral images processed to isolate tomato canopy reflectance. Our findings revealed that the earliest growth stage at which broomrape could be detected with acceptable Accuracy was at 897 GDD, achieving an overall Accuracy of 79.09% and a recall rate for broomrape of 70.36%, without the integration of subsequent growing stages. However, when considering sequential growing stages, the LSTM models applied across four distinct scenarios with and without SMOTE augmentation indicated significant improvements in the identification of broomrape-infested plants. The best-performing scenario, which integrated all growth stages, achieved an overall Accuracy of 88.37% and a Recall rate of 95.37%. These results demonstrate the LSTM network's robust potential for early broomrape detection and highlight the need for further data collection to enhance the model's practical application. Looking ahead, the study's approach promises to evolve into a valuable tool for precision agriculture, potentially revolutionizing the management of crop diseases and supporting sustainable farming practices.

## 1. INTRODUCTION

California's agriculture industry heavily relies on tomato production, with the state producing over 90% of the nation's supply. However, this vital sector faces a significant threat from branched broomrape (*Phelipanche ramosa* (L.) Pomel), a parasitic plant that can severely impact crop yields. Lacking chlorophyll, branched broomrape is a stealthy parasite that spends most of its life cycle underground, attaching to the roots of host plants to extract water and essential nutrients. This underground lifestyle allows the parasite to evade early detection, as visible symptoms on the host plants appear only after substantial damage has been inflicted[1], [2]. The lifecycle of branched broomrape involves complex interactions with its host, making it particularly challenging to manage. After germination, triggered by stimulants exuded by host plant roots, broomrape seeds develop a haustorium—a specialized organ that penetrates the host roots to siphon off their nutrients. As

---

a result, infested tomato plants often exhibit stunted growth wilting and can eventually die if the infestation is severe. This parasitism leads to considerable economic losses due to reduced yield and quality of the tomatoes [3].

Traditional methods for managing broomrape involve widespread chemical applications that are costly and pose environmental and health risks. The herbicides currently used are often non-selective, targeting not only broomrape but also beneficial organisms and nearby crops. For example, the use of glyphosate and other systemic herbicides has been prevalent; however, their effectiveness varies and often requires high doses that can be detrimental to the surrounding ecosystem [4], [5]. Moreover, the reliance on chemical treatments highlights the critical need for more precise and sustainable management strategies. Pesticides such as sulfosulfuron and imazapic have shown some efficacy in controlling broomrape. Nevertheless, these solutions also risk harming non-target plant species and contributing to the development of herbicide resistance [1]. The environmental impact, coupled with the potential for reduced effectiveness over time, underscores the limitations of conventional detection and control methods. Recent studies advocate for the development of targeted solutions that minimize unnecessary chemical usage while effectively controlling infestations. Innovations in early detection technologies and integrated pest management (IPM) strategies are being explored to reduce reliance on broad-spectrum herbicides. These approaches aim to achieve a more sustainable balance, using chemicals only when and where they are truly needed, thus preserving the agricultural environment and reducing the overall ecological footprint of crop production [6].

In response to these challenges, there has been a growing interest in leveraging advanced remote sensing technologies. Research has demonstrated that hyperspectral and multispectral imaging, which capture not only visible light but also infrared and thermal bands, can significantly enhance the detection of physiological changes in plants caused by diseases or pests. These technologies are particularly effective in identifying early signs of broomrape infestation, which is crucial for implementing timely and targeted treatment strategies. The application of drones equipped with multispectral sensors offers a promising avenue for precisely mapping broomrape infestations within crop fields. By detecting subtle differences in reflectance spectra emitted from infested versus healthy plants, these tools enable farmers to apply herbicides selectively rather than over entire fields. This targeted approach conserves resources, minimizes environmental impact, and reduces herbicide resistance risk [7]. Furthermore, the complexity and volume of data generated by these remote sensing technologies require sophisticated analytical approaches. Recent advancements in machine learning and deep learning, particularly the use of Long Short-Term Memory (LSTM) networks, have been identified as effective in modeling temporal changes in plant health, which are crucial for detecting infestations before they become visually apparent. Moreover, addressing the challenge of data imbalance—where healthy plants significantly outnumber diseased ones—has led researchers to adopt techniques such as the Synthetic Minority Over-sampling Technique (SMOTE). This approach helps to balance datasets, thereby improving the accuracy of predictive models aimed at identifying disease presence [8].

The overarching goal of this research is to develop a robust detection system for branched broomrape in tomato farms using drone-captured multispectral imagery. We aim to achieve this through three primary objectives: evaluating the effectiveness of drone-based multispectral imagery in detecting spectral changes in the canopy and leaves of tomato plants that may indicate broomrape infestation despite the parasite's predominantly subterranean life cycle; determining the earliest growing degree day at which broomrape can be detected with acceptable Accuracy using our deep learning model; and assessing how the incorporation of sequential data reflecting different growth stages into a Long Short-Term Memory (LSTM) model influences the Accuracy of broomrape detection. This innovative approach seeks to mitigate the impact of broomrape on tomato yields and establish a model for sustainable pest management practices in agriculture. By focusing on the integration of technology and modeling, this research intends to provide actionable insights that enhance the precision and effectiveness of broomrape detection methods.

## 2. METHODOLOGY

### 2.1. DATA COLLECTION AND SENSOR

The data collection for this study was conducted on a tomato farm located in Woodland, Yolo County, California, US. This farm was specifically selected due to its history of branched broomrape infestations over the past few years, making it an ideal site for examining the effectiveness of remote sensing in detecting this parasite. To align our data collection with the key developmental stages of tomatoes, we organized the collection around the concept of growing degree days (GDD) [9]. The GDD is calculated according to Equation 1.

$$\text{GDD} = \frac{T_{max} + T_{min}}{2} - T_{base} \tag{1}$$

where $T_{base}$ for tomatoes is set at 10 degrees Celsius. We calculated GDD daily, utilizing local meteorological data to obtain $T_{min}$ and $T_{max}$. The specific GDDs targeted for data collection were 324, 574, 897, 1195, and 1556 GDD.

For ground-truthing, we marked and monitored 300 random tomato plants throughout the growing season. These plants were regularly scouted for signs of broomrape emergence. By the end of the season, we identified 49 plants infected with broomrape and 251 healthy plants. This ground truth data was essential for validating the remote sensing data collected via aerial surveys.

Aerial data was captured using a DJI Matrice 210 drone (DJI Matrice 210 Drone, SZ DJI Technology Co., Ltd., Shenzhen, China), which is capable of a maximum takeoff weight of 6.14 kg, allowing it to carry the MicaSense Altum-PT sensor (MicaSense Altum-PT Sensor, MicaSense Inc., Seattle, USA). This sensor provided comprehensive spectral and thermal data, which is crucial for our analysis. The Altum-PT sensor features include a resolution of 2064 x 1544 (3.2MP) for each multispectral band and 4112 x 3008 (12MP) for the panchromatic band, covering spectral bands such as Blue (475nm ±32nm), Green (560nm ±27nm), Red (668nm ±14nm), Red Edge (717nm ±12nm), and NIR (842nm ±57nm). Additionally, it includes a thermal infrared band (FLIR LWIR) spanning 7.5-13.5um, which is radiometrically calibrated. The ground sampling distance (GSD) for the multispectral bands was 5.28 cm per pixel at a flight altitude of 120 meters, while the thermal GSD was 33.5 cm per pixel, and the panchromatic GSD was 2.49 cm per pixel at the same altitude.

Figure 1 in our study illustrates the map of California counties with the most tomato farms as reported in the USDA cropland database. We utilized ArcGIS Pro to generate this map, and the specific location of our target farm is indicated on this map, providing a geographic context for our data collection efforts. This figure includes detailed images of both the farm's layout and a close-up view of a tomato plant flagged as infected by broomrape during field scouting. These elements collectively enhance our understanding of the geographic and agricultural settings pertinent to our study.

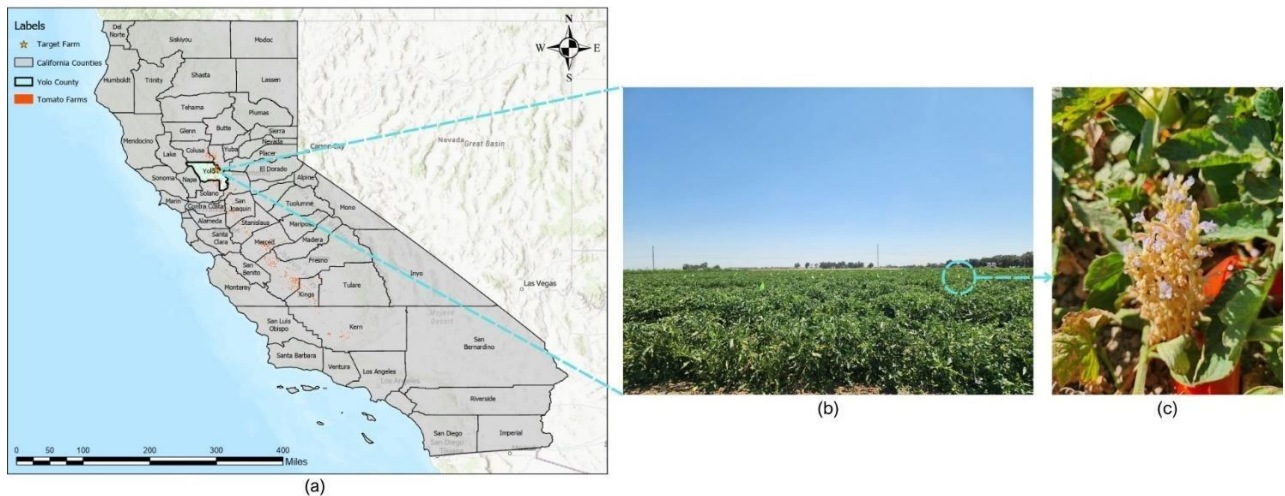

Figure 1: Comprehensive visual overview of the study area: (a) Map of California showing major tomato farming counties and the location of our target farm, generated with ArcGIS Pro; (b) Front view of the target tomato farm; (c) Close-up of a tomato plant flagged as infected by broomrape.

## 2.2. AERIAL MULTISPECTRAL DATA PROCESSING

The initial output from the Altum-PT sensor was provided as digital numbers. For accurate analysis, these needed to be converted into reflectance values to correct for light conditions and sensor differences. We accomplished this using reflectance panels placed within the farm, which served as calibration targets. Additionally, a Python script was developed to automate the conversion process, ensuring consistency and precision in transforming digital numbers to reflectance data across all captured images. Although the aerial data covered the entire farm scene, our focus was on the 300 tomato plants previously flagged for detailed analysis. We cropped the multispectral images to isolate these plants from the broader field data to encompass only these specific areas. This step was critical in narrowing our dataset for more focused and effective processing.

Given that the cropped images contained both plant canopy and soil, it was essential to segment out the canopy for accurate reflectance analysis. We employed the Soil-Adjusted Vegetation Index (SAVI) to minimize the influence of soil brightness, which is particularly important in our settings where the vegetative cover was not uniformly dense [10]. The SAVI is calculated according to Equation 2.

$$SAVI = \frac{(NIR - RED)}{(NIR + RED + L)} \times (1 + L) \qquad (2)$$

where *NIR* represents the pixel values from the near-infrared band, *Red* denotes the pixel values from the red band, and *L* is a factor adjusted for green vegetative cover, set at 0.5 due to the approximately equal distribution of canopy and soil in our images. This adjustment was particularly suited to our farm's conditions, facilitating effective discrimination of plant material from the soil background. The threshold for masking the canopy was set at 0.5, optimizing the separation of vegetative and non-vegetative elements within the images.

Figure 2 in our study illustrates the application of these methodologies, showing images of the entire tomato farm captured across all seven spectral bands. It also includes samples of the cropped tomato plants, both before and after applying the SAVI index and the subsequent masking of the canopy from the soil. This visualization demonstrates the effectiveness of our data processing approach and provides a clear comparison between raw and processed images.

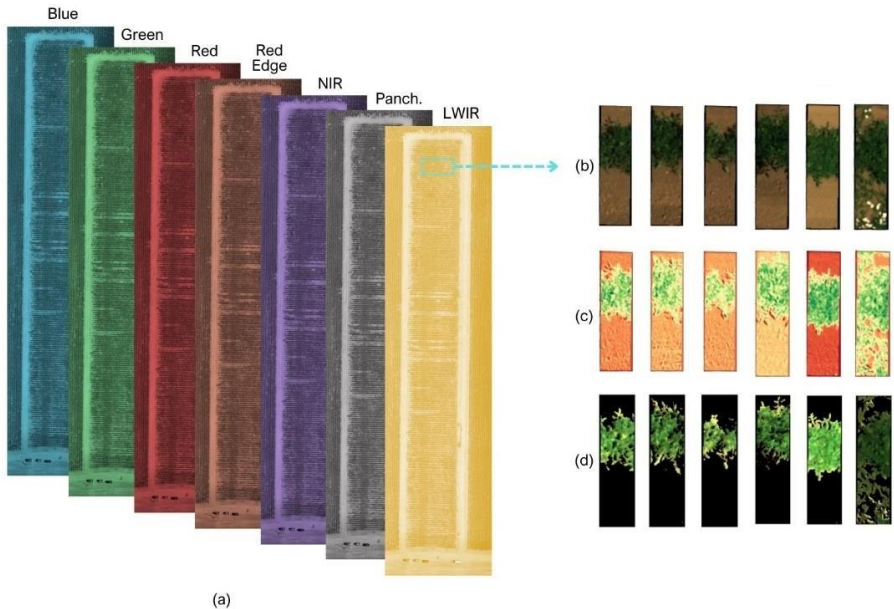

Figure 2: Visualization of data multispectral processing steps: (a) shows the entire spectral range across seven bands, (b) displays cropped RGB visualization of tomato plants, (c) illustrates the application of the SAVI index, and (d) depicts the masking of the canopy from the soil.

### 2.3. FEATURE EXTRACTION AND DATA AUGMENTATION

In the process of preparing our data for deep learning analysis, we first extracted significant statistical features from the segmented tomato canopy images, which contained multiple pixels per plant. As depicted in Figure 3, we analyzed histogram samples from each spectral band across all growing degree days, comparing healthy and infected tomato plants. An essential rationale for employing histogram feature extraction stems from the varying spatial resolutions across the seven spectral bands used in our study. Direct usage of images with differing resolutions can lead to inconsistencies in data interpretation. By extracting histogram features, we ensure that our model is guided by the most pertinent features among the pixels of each canopy, enhancing data clarity and reducing noise. This approach facilitates more efficient processing and significantly accelerates the overall data handling and analysis. Following the method outlined by [11], we extracted key features such as mean, standard deviation, third moment, uniformity, entropy, and gray level range. These features capture the statistical essence of pixel intensity distributions across spectral bands, enabling us to distinguish between healthy and infected plants effectively. The equations for calculating these parameters are detailed in Table 1, ensuring a consistent approach across all growing degree days for both plant conditions.

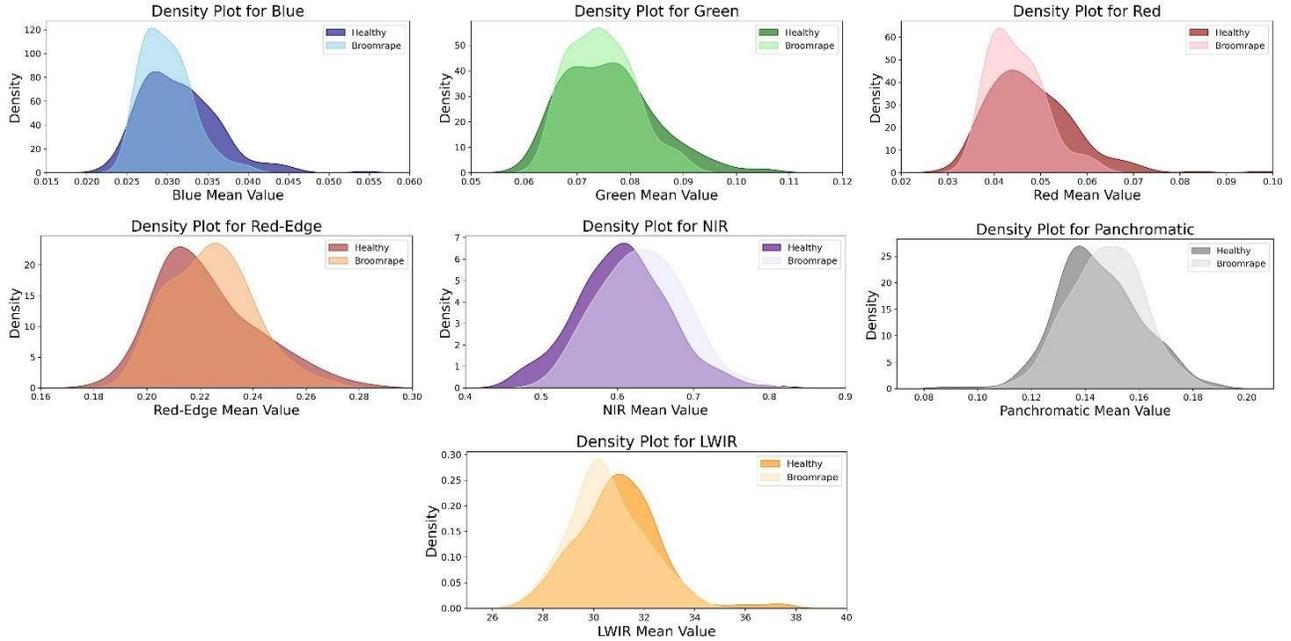

Figure 3: KDE plots of spectral band data at 897 GDD for healthy versus infected tomato plants, showcasing the distribution differences in reflectance and temperature patterns.

Table 1. Features extracted from all seven bands

| Feature | Equation |
| --- | --- |
| Mean | $\mu = \sum_i p(i)$ |
| Standard deviation | $\sigma = \sqrt{\sum_i (i - \mu)^2 \, p(i)}$ |
| Smoothness | $1 - 1/(1 + \sigma^2)$ |
| Third moment | $\sum_i (i - \mu)^3 \, P(i)$ |
| Uniformity | $\sum_i P(i)^2$ |
| Entropy | $-\sum_i P(i) \log \{p(i)\}$ |
| Gray level range | $\max\{i | p(i) \neq 0\} - \min(i | p(i) \neq 0)$ |

Addressing the challenge posed by the imbalanced dataset resulting from the broomrape lifecycle was crucial. At the season's end, we observed that only 49 of the 300 targeted plants were infected with broomrape, while the remaining 251 were healthy. To manage this imbalance, we implemented two distinct approaches. The first approach involved selecting all 49 infected plants and an equal number of healthy plants whose features most closely matched the overall mean of the healthy class. This method helped balance the dataset but resulted in the loss of a significant number of healthy samples. Our second strategy was to augment the data for the infected class using the Synthetic Minority Over-sampling Technique (SMOTE), as described by [8]. This technique synthetically generates new examples from the minority class, thus increasing the infected sample size to match the 251 healthy plants. SMOTE enhances the training dataset's diversity and representativeness, enabling more robust model training and reducing the risk of overfitting to the majority class. These methodologies facilitated a balanced approach to data analysis and ensured that our deep learning models were trained on representative and comprehensive datasets, improving their Accuracy and generalizability in detecting broomrape-infected plants.

### 2.4. DEEP LEARNING MODEL AND EVALUATION METRICS

Our deep learning model was developed using the Python programming language, with computations facilitated by Google Colab's robust environment that provides access to NVIDIA Tesla T4 GPUs with 16GB of VRAM. This setup was chosen to leverage the extensive computational resources necessary for training deep learning models. The core of our model is a Long Short-Term Memory (LSTM) network, specifically chosen for its proficiency in handling time series data. LSTM networks are ideal for agricultural applications where the temporal sequence of observations is crucial, as they can maintain information over long intervals—essential for capturing the developmental patterns of tomato plants across various growth stages [12].

The LSTM model architecture included two LSTM layers, each followed by dropout layers to mitigate overfitting. L2 regularization was applied to the network to prevent the coefficients from reaching large values, which helps reduce overfitting and improve model generalization. Kernel constraints were also used to regulate the magnitude of the weights. The model comprised a total of 42,689 parameters, all of which were trainable, with no non-trainable parameters. This setup ensures that our model can adapt flexibly to the nuances of the input data. The model architecture was designed to output binary classifications, indicating the health status of the tomato plants as either healthy or infected by broomrape. The architecture's configuration aimed at optimizing the detection Accuracy by fine-tuning the balance between sensitivity and specificity.

Figure 4 in our study presents a simplified diagram of our LSTM model architecture, created using Netron (https://netron.app), an open-source tool for visualizing neural network models. This diagram effectively illustrates the layers and their interconnections, facilitating a clear understanding of the data flow and processing within the model. The visual representation provided by Netron helps highlight the structural complexities of our LSTM network in a user-friendly manner, making it easier to comprehend the operational dynamics of the model.

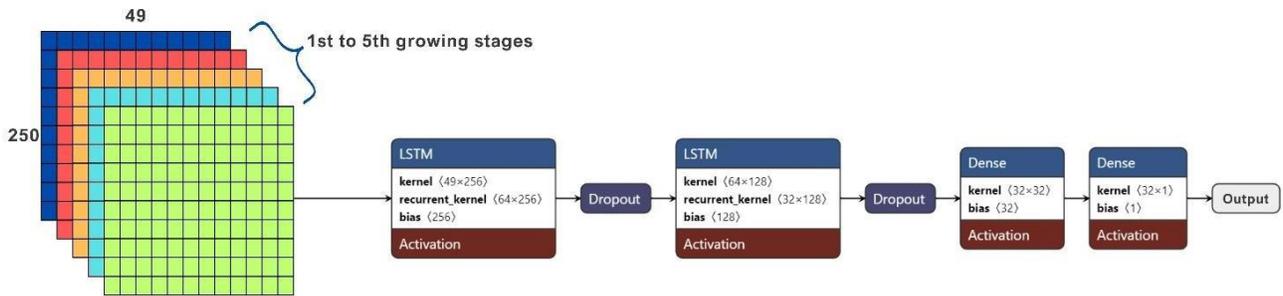

Figure 4: Simplified diagram of the LSTM model architecture, illustrating the arrangement and interconnections of various layers.

To rigorously evaluate the robustness and adaptability of our LSTM model, we implemented it across four distinct scenarios. In the first scenario, the model was applied separately to the dataset from each growth stage before any data augmentation, treating data from each stage as a unique dataset. The second scenario extended this by employing the LSTM in a time-series manner before data augmentation, where time served as a third dimension. Here, training commenced with data from the first growth stage and was sequentially expanded to include data from subsequent stages through to the fifth. In the third scenario, similar to the first, the LSTM was again applied to each stage separately but followed the application of SMOTE to augment the data and achieve class balance. The fourth scenario mirrored the second; however, it incorporated SMOTE prior to model training to ensure a balanced dataset from the outset. These scenarios were crafted to rigorously test the LSTM's capability under varied conditions, emphasizing its potential utility in real-world agricultural settings where early detection of diseases is critical for effective management and mitigation. To comprehensively assess the performance of our LSTM model, we employed several key metrics traditionally used in binary classification tasks, including Recall, Precision, F1-Score, and Overall Accuracy. Recall, or Sensitivity, measures the model's ability to identify all actual positive cases (infected plants) correctly and is calculated as:

$$Recall = \frac{TP}{TP + FN} \quad (3)$$

where *TP* represents true positives, and *FN* represents false negatives. Precision assesses the Accuracy of the positive predictions made by the model, defined as:

$$Precision = \frac{TP}{TP + FP} \quad (4)$$

where *FP* denotes false positives. The F1-Score, which is the harmonic mean of Precision and Recall, provides a balance between these metrics and is calculated using:

$$F1\ Score = 2 \times \frac{Precision \times Recall}{Precision + Recall} \quad (5)$$

This score is especially useful in uneven class distribution scenarios, ensuring that the model's performance is not biased toward the majority class. Lastly, Overall Accuracy provides a straightforward indication of the model's effectiveness across all classifications and is formulated as follows:

$$Overall\ Accuracy = \frac{TP + TN}{TP + TN + FP + FN} \quad (6)$$

Where *TN* stands for true negative. These metrics offer a multi-faceted view of the model's capabilities. Recall ensures that the model detects as many infected plants as possible, Precision minimizes the false alarms, and the F1-Score provides a single measure reflecting the balance between Recall and Precision. Overall Accuracy gives a holistic view of the model's total effectiveness in correctly classifying both positive and negative samples. The rigorous application and interpretation of these metrics follow the guidelines and recommendations discussed by [13], who underscore the importance of these measures in evaluating the performance of classification models.

## 3. RESULTS AND DISCUSSION

Our investigation into the effectiveness of Long Short-Term Memory (LSTM) networks for the detection of broomrape-infested tomato plants across varying growing degree days (GDDs) and scenarios revealed insightful findings. Table 2 presents a precise evaluation of our LSTM model's performance on the test set, comprising 20% of our data, with training and validation sets accounting for 65% and 15%, respectively.

Table 2. Performance Metrics of LSTM Model Across Different Scenarios and Growing Degree Days on Test Set

| | GDDs | Augmentation | Metrics (%) | | | | | | |
|---|---|---|---|---|---|---|---|---|---|
| | | | Broomrape Class | | | Healthy Class | | | Overall |
| | | | P; | R; | F; | P; | R; | F; | Accuracy |
| Scenario1 | 324 | None | 50.14 | 10.22 | 17.13 | 47.34 | 89.21 | 62.48 | 47.37 |
| | 574 | | 80.12 | 40.46 | 53.37 | 57.13 | 89.18 | 70.21 | 63.16 |
| | 897 | | 60.49 | 60.34 | 60.24 | 56.42 | 56.31 | 56.27 | 57.89 |
| | 1195 | | 30.44 | 30.19 | 30.38 | 22.16 | 22.12 | 22.29 | 26.32 |
| | 1556 | | 70.14 | 70.36 | 70.49 | 67.16 | 67.24 | 67.47 | 68.42 |
| Scenario2 | 324, 574 | None | 75.26 | 30.47 | 43.28 | 53.12 | **89.34** | 67.41 | 57.89 |
| | 324. 574, 897 | | 75.16 | 60.32 | 67.16 | 64.23 | 78.42 | 72.47 | 68.42 |
| | 324, 574, 897, 1195 | | 54.37 | 70.46 | 61.31 | 50.12 | 33.47 | 40.16 | 52.63 |
| | 324, 574, 897, 1195, 1556 | | 62.39 | 50.25 | 56.45 | 55.12 | 67.37 | 60.41 | 57.89 |
| Scenario3 | 324 | SMOTE | 81.19 | 70.05 | 75.16 | 73.26 | 84.09 | 78.12 | 76.74 |
| | 574 | | 63.02 | 51.16 | 56.49 | 59.06 | 70.01 | 64.26 | 60.47 |
| | 897 | | 86.41 | 70.36 | 77.16 | 75.21 | 88.35 | 81.19 | 79.09 |
| | 1195 | | 55.42 | 49.05 | 52.46 | 54.36 | 60.48 | 57.44 | 54.65 |
| | 1556 | | 82.01 | 72.12 | 77.26 | 75.41 | 84.03 | 79.34 | 77.91 |
| Scenario4 | 324, 574 | SMOTE | 85.16 | 67.37 | 75.44 | 73.08 | 88.41 | 80.35 | 77.91 |
| | 324, 574, 897 | | 85.16 | 67.44 | 75.06 | 73.49 | 88.34 | 80.21 | 77.91 |
| | 324, 574, 897, 1195 | | **88.16** | 84.22 | 86.37 | 84.34 | 88.09 | 86.11 | 86.06 |
| | 324, 574, 897, 1195, 1556 | | 84.46 | **95.37** | **89.46** | **95.12** | 81.06 | **88.16** | **88.37** |

In Scenario 1, where the dataset was not subjected to augmentation, and each class comprised only 49 samples, the model performance was found to be suboptimal. This scenario underscores the challenge of limited data in a high-dimensional feature space, which can lead to poor generalization and low predictive performance. Scenario 2, which introduced time as an additional dimension but maintained the small sample size, exhibited negligible improvements in performance metrics. This outcome is a manifestation of the curse of dimensionality, where an increase in features without a corresponding increase in data volume can hinder the model's ability to learn effectively.

A significant enhancement in model performance was observed in Scenario 3, which utilized SMOTE for data augmentation, resulting in 251 samples for each class and processing each growth stage independently. This augmentation dramatically increased classification Accuracy, demonstrating the effectiveness of balancing class distribution in training datasets. In this scenario, we found that the earliest growth stage at which broomrape could be detected with acceptable Accuracy was at 897 GDD, achieving an overall Accuracy of 79.09% and a recall rate for broomrape of 70.36% without the integration of subsequent growing stages. Finally, Scenario 4 emerged as the most successful approach, incorporating SMOTE data augmentation and integrating time as the third dimension across all five growth stages. This comprehensive strategy yielded an overall Accuracy of 88.37%, with a remarkable Recall rate of 95.37% for the broomrape class. This high Recall rate is particularly noteworthy as it reflects the model's capability to identify the majority of infected plants, which is the primary objective in practical applications for early detection.

The results suggest that employing SMOTE to achieve a balanced dataset, coupled with including temporal growth stages in the model, significantly enhances LSTM performance. It improved the overall Accuracy and boosted precision, recall, and the F1 score across classes. Such improvements are critical for deploying deep learning models in precision agriculture, where the early and accurate detection of plant diseases can lead to more targeted interventions and better crop management. Moreover, the influence of SMOTE data augmentation and the temporal inclusion of growth stages are visually represented in Figure 5. This figure elucidates the overall Accuracy across various scenarios, contrasting the impacts of both with and without the augmentation and the inclusion of time as a third dimension, offering a comprehensive insight into the performance enhancements enabled by these methodological enhancements.

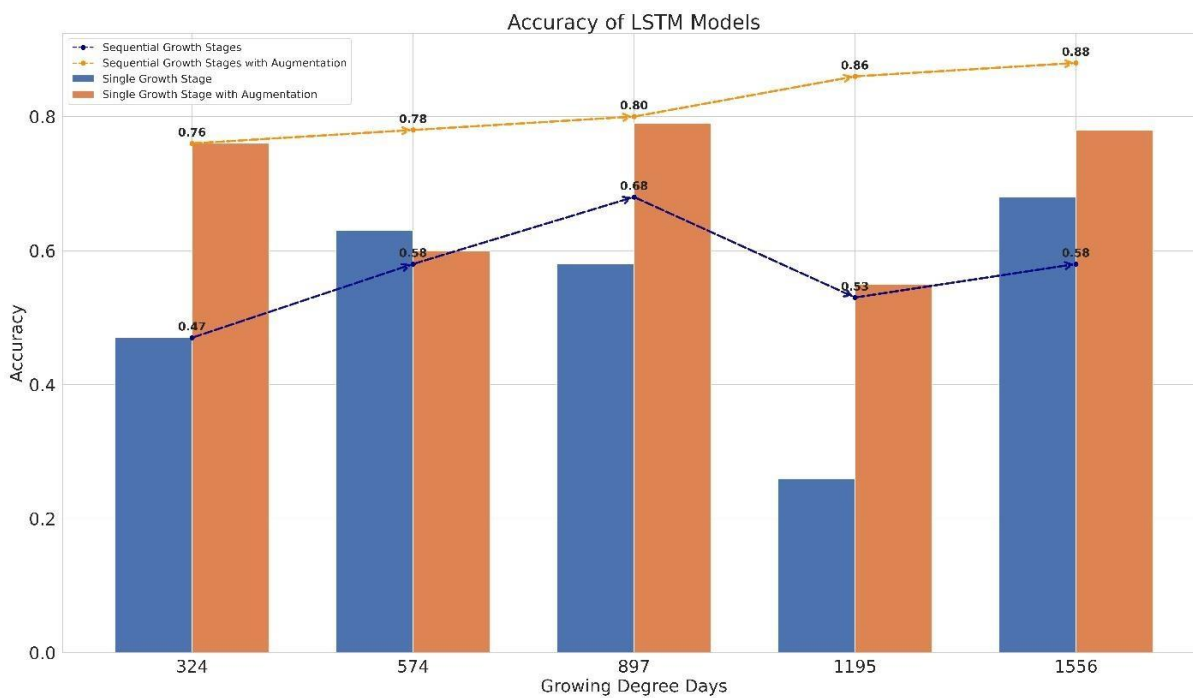

Figure 5. Graphical representation of overall Accuracy across the four scenarios, comparing the impact of SMOTE data augmentation and the incorporation of temporal growth stages.

Delving into the details of Scenario 4, which integrates all growth stages and employs SMOTE augmentation, we observe the most promising results. As depicted in Figure 6, panels (a) illustrate the model's loss and Accuracy over 100 epochs, respectively. The convergence of these metrics for both training and validation sets indicates that the model has effectively navigated the challenges of underfitting, achieving a robust fit to the data.

The kernel density estimate (KDE) versus probability plot in panel (b) of Figure 6 provides further insight into the model's classification prowess. This plot distinctly shows that the samples are well-separated by the classification threshold of 0.5, with the majority of infected plant predictions amassing near 0 and healthy plant predictions clustering closer to 1. The clear distance of the sample probabilities from the threshold underscores the model's decisive classification ability, particularly for the infected class, affirming the minimal risk of marginal or random prediction errors. Panel (c) features the confusion matrix for our model's predictions, offering a straightforward visualization of the true positive and negative rates, as well as the instances of misclassification. The confusion matrix reveals that out of 43 tomato plants identified as infected by broomrape, 41 were accurately predicted, underscoring the model's precision and its substantial potential in practical applications for early disease detection.

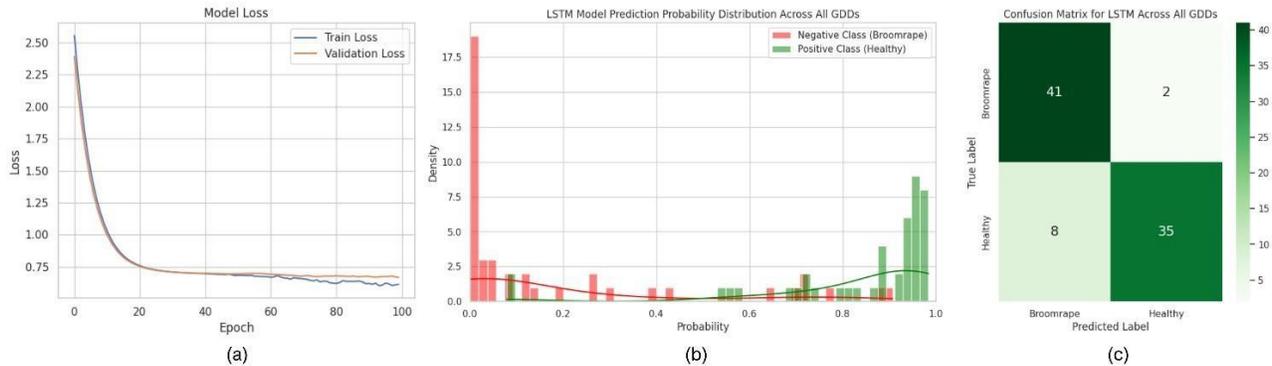

Figure 6: LSTM model evaluation for Scenario 4 with (a) training and validation loss over 100 epochs, (b) KDE versus probability plot of model predictions, and (c) confusion matrix, demonstrating high predictive Accuracy.

While the LSTM model's performance in a sequential manner—incorporating all growth stages—has proven to be highly effective, particularly on an unseen test set, the reliance on synthetic augmentation like SMOTE, coupled with the extensive feature set across multiple spectral bands, highlights inherent challenges. These challenges underscore the need for more real-world data in both classes to train the model adequately. Collecting ample real-world data to balance the dataset naturally presents logistical difficulties, as the detection of broomrape infestation prior to the end of the growing season is currently not feasible. This limitation presents a barrier to obtaining a sufficiently large and balanced dataset that is both cost-effective and less labor-intensive. In practice, while the model shows considerable promise, the ideal path forward would involve extensive data collection across diverse farms and conditions to address the issue of imbalanced datasets organically. This approach would add depth and variability to the training data and enhance the model's ability to generalize across different scenarios. Additionally, the deployment of the model in subsequent years and its continuous refinement with new data would contribute to a more robust tool—one that farmers could eventually employ for early broomrape detection.

## 4. CONCLUSION

This research set out to confront the challenges posed by branched broomrape (Phelipanche ramosa) to California's tomato production, a significant agricultural sector of the state. The clandestine nature of broomrape's lifecycle, causing severe damage from beneath the soil, necessitated a novel approach to early detection. Through the use of drone-captured multispectral imagery and the implementation of LSTM networks enhanced by SMOTE for data augmentation, we established a method capable of identifying infected plants with a high degree of Accuracy. Our results showcased the strength of combining temporal data with balanced datasets, particularly in Scenario 4, which yielded an overall Accuracy of 88.37% and a Recall rate of 95.37% for detecting broomrape, signaling a substantial step forward in early disease intervention. The findings align with our primary objectives by confirming that multispectral imagery is effective in detecting spectral changes indicative of broomrape infestation. Additionally, the earliest detectable stage with acceptable Accuracy was identified at 897 GDD, highlighting the potential for timely intervention. The integration of all growth stages into the LSTM model significantly enhanced detection Accuracy, demonstrating the value of sequential data analysis in this context. However, the research also highlighted the intrinsic limitations of synthetic data augmentation and the need for more comprehensive real-world data to train and validate the model. Looking ahead, extensive data collection across a variety of farms and conditions will be vital for fine-tuning the model's efficacy. Our work paves the way for future developments, aiming to evolve our LSTM model into a robust, field-ready tool for farmers to manage and mitigate the effects of broomrape more efficiently and sustainably. This commitment to innovation in precision agriculture holds the promise of bolstering crop management and securing the future of tomato production in California and beyond.

## 5. ACKNOWLEDGMENT

The authors would like to gratefully acknowledge the funding from the California Tomato Research Institute (CTRI). Their contribution has been essential to our research into combating broomrape in tomato crops.